\documentclass[pdflatex,sn-mathphys-ay]{sn-jnl}% Math and Physical Sciences Author Year Reference Style

\makeatletter
\@twosidefalse
\@mparswitchfalse
\makeatother

\usepackage{graphicx}%
\usepackage{multirow}%
\usepackage{amsmath,amssymb,amsfonts}%
\usepackage{amsthm}%
\usepackage{mathrsfs}%
\usepackage[title]{appendix}%
\usepackage{xcolor}%
\usepackage{textcomp}%
\usepackage{manyfoot}%
\usepackage{booktabs}%
\usepackage{algorithm}%
\usepackage{algorithmicx}%
\usepackage{algpseudocode}%
\usepackage{listings}%
\usepackage{tabularx}

\PassOptionsToPackage{hyphens}{url}
\usepackage{hyperref}

\usepackage{etoolbox}
\apptocmd{\sloppy}{\hbadness 10000\relax}{}{}

\theoremstyle{thmstyleone}%
\theoremstyle{thmstyletwo}%

\theoremstyle{thmstylethree}%

\begin{document}

\title[Article Title]{Chatbot Conversations in Physics Education: Using Artificial Intelligence to Analyze Student Reasoning through Computational Grounded Theory}

\author[1, 2]{\fnm{Atharva} \sur{Dange}}\email{dange98@mit.edu}

\author[1]{\fnm{Ramon E.} \sur{Lopez}}\email{relopez@uta.edu}

\affil[1]{\orgdiv{Physics}, \orgname{University of Texas at Arlington}, \orgaddress{\city{Arlington}, \postcode{76019}, \state{}, \country{USA}}}

\affil[2]{\orgdiv{Physics}, \orgname{Massachusetts Institute of Technology}, \orgaddress{\city{Cambridge}, \postcode{02139}, \state{}, \country{USA}}}

\abstract{This study applies Computational Grounded Theory (CGT) to analyze student misconceptions using interaction data from an AI-powered chatbot deployed in a university-level Modern Physics course. The chatbot — the \textit{UTA Study Buddy Bot} — engaged students in peer-like problem-solving conversations throughout the semester, generating a rich dataset of over 10 million tokens. To explore patterns in student reasoning and identify recurring conceptual difficulties, we implemented a CGT pipeline that combined natural language processing, unsupervised clustering of sentence-level vector embeddings, human interpretation of emergent themes, and supervised learning to evaluate the generalizability of identified categories. Preliminary results revealed persistent misconceptions in areas such as relativistic momentum and quantum energy levels, along with distinctive trends in how students phrased their questions and expressed uncertainty. These findings underscore the potential of CGT as a scalable, theory-aligned approach for extracting insights from chatbot dialogues and guiding the development of more adaptive, AI-driven educational tools in physics instruction.}

\keywords{Chatbots, Computational Grounded Theory, Generative AI, Student Misconceptions, LLMs, Physics Education}

%%\pacs[JEL Classification]{D8, H51}

%%\pacs[MSC Classification]{35A01, 65L10, 65L12, 65L20, 65L70}

\maketitle

\section{Introduction}\label{sec1}

Artificial intelligence (AI) is rapidly transforming educational environments, offering new modes of engagement and support through tools such as chatbots, intelligent tutoring systems, and large language models \citep{adiguzel_revolutionizing_2023, alqahtani_emergent_2023, sharma_role_2025}. In the context of physics education, these technologies provide students with immediate, conversational access to assistance on homework, exam preparation, and conceptual understanding. While these interactions often mimic peer discussion or one-on-one tutoring, they also produce extensive textual data that remains underutilized in education research. Such dialogue logs may reveal how students articulate their uncertainties, pose conceptual questions, and exhibit patterns of reasoning that reflect deeper misconceptions \citep{odden_using_2024}. 

Qualitative methods in physics education research (PER) \citep{otero_qualitative_2023} have historically enabled rich insights into student thinking, particularly around difficult or abstract topics like quantum mechanics, relativity, and thermodynamics \citep{bungum_quantum_2018, forbus_using_1998}. However, these methods, which typically rely on interviews, written responses, or small-group observations, are time-intensive and do not scale easily to large datasets. As educational tools increasingly generate massive amounts of unstructured textual data, traditional qualitative approaches struggle to capture their full potential \citep{wiedemann_text_2016}. The result is a growing gap between the richness of available student discourse and the feasibility of analyzing it meaningfully. 

Student misconceptions have long been a central focus of PER \citep{kaltakci_gurel_review_2015}, especially in areas such as relativistic dynamics and quantum phenomena \citep{henriksen_relativity_2014, leonardi_analysis_2024, styer_common_1996}. These misconceptions often stem from deeply held intuitive models that resist correction, even after direct instruction \citep{mcdermott_millikan_1991, nesher_towards_1987}. Traditional assessments may fail to capture how students genuinely conceptualize these topics or how they frame their difficulties \citep{wiggins_true_2011}. By analyzing informal student expressions in chatbot conversations, this study aims to capture both the content and the form of their misconceptions, providing a more nuanced picture of learning-in-action.

\subsection{Computational Grounded Theory}\label{subsec1.1}

Computational Grounded Theory (CGT) \citep{carlsen_computational_2022, fang_topic_2023} offers a promising bridge between large-scale data and theory-driven qualitative analysis. Originally proposed by Nelson \citep{nelson_computational_2020}, CGT combines unsupervised machine learning (ML) and natural language processing techniques with the iterative, interpretive coding practices of grounded theory \citep{salam_integrating_2019, sheetal_machine_2024}. The method unfolds in three steps: pattern detection using computational tools \citep{qiu_pre-trained_2020}, pattern refinement through deep human reading \citep{nelson_computational_2020}, and pattern confirmation via further analysis or modeling \citep{tschisgale_integrating_2023}. This hybrid approach maintains the contextual and theoretical depth of traditional qualitative research while addressing issues of scalability, reproducibility, and efficiency \citep{lindberg_aron_developing_2020}. 

This paper applies CGT to analyze a semester-long dataset of student-chatbot conversations in a modern physics course at the University of Texas at Arlington. The course was online synchronous, and students had the option to use an AI-powered assistant—the \textit{UTA Study Buddy Bot}—to seek clarification on course topics, problem-solving strategies, and exam preparation. They also extensively used the chatbot to solve homework, practice exam questions and clear core physics concepts. These interactions produced a corpus of over 10 million tokens \citep{grefenstette_what_1994}. Instead of relying on structured diagnostics, this study explores spontaneous, natural language student inquiries as a window into their evolving understanding and misconceptions in physics \citep{aguilar_design_2022, lieb_student_2024, trout_artificial_2025}. 

\begin{figure}[h]
\centering
\includegraphics[width=\textwidth]{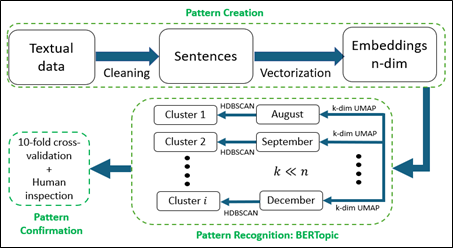}
\vspace{-1em} % <-- optional if caption is too far
\caption{CGT Pipeline for Clustering Student Reasoning via BERTopic}
\label{Fig. 1}
\end{figure}

The core of the CGT pipeline involves dimensionality reduction (e.g., via UMAP) \citep{allaoui_considerably_2020, becht_dimensionality_2019, mcinnes_umap_2018} and clustering algorithms (e.g., HDBSCAN, SNN, DENCLUE), which group semantically similar sentences into distinct clusters \citep{calio_advancing_2022, khader_new_2020, mehta_comprehensive_2024}. Each cluster corresponds to student messages from a particular module of the course, enabling fine-grained tracking of conceptual themes across different topics (Modules 1–13). The rightward blue arrows labeled “Pattern Recognition” represent this machine-driven detection of recurring linguistic patterns in student discourse. Importantly, the pipeline then loops back through human inspection, where researchers interpret these clusters and validate or refine the themes they represent—a process labeled “Pattern Confirmation”. This hybrid method leverages the scale and consistency of machine learning while retaining the interpretive depth of grounded qualitative analysis. See Fig. \ref{Fig. 1} for the graphical representation of this framework.

\begin{figure}[h]
\centering
\includegraphics[width=\textwidth]{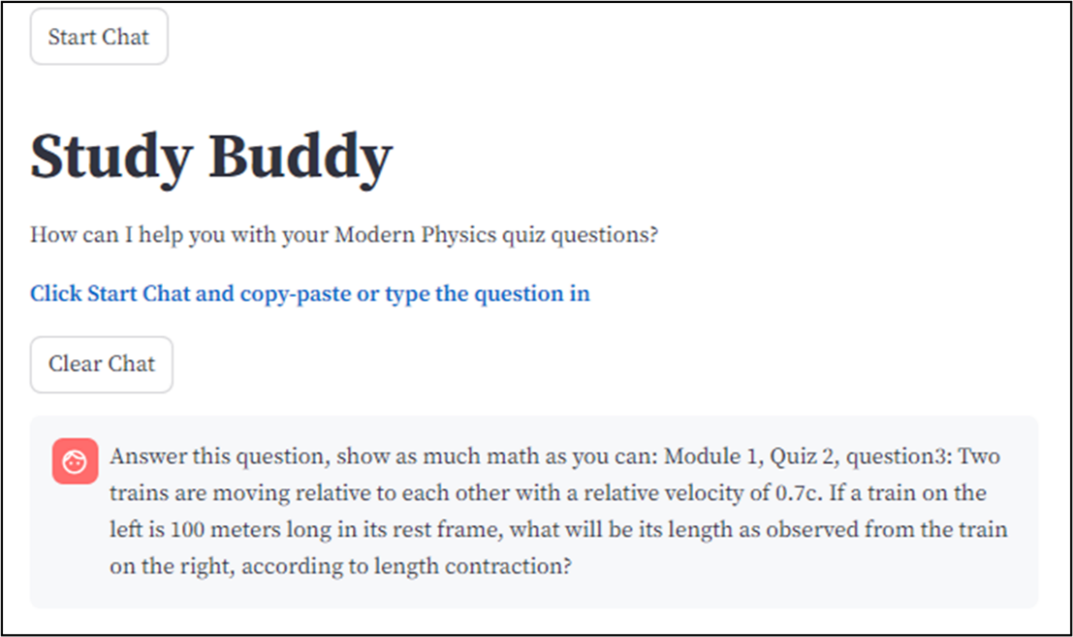}
\vspace{-1em} % <-- optional if caption is too far
\caption{Example of a typed question within the UTA Study Buddy Bot interface}
\label{Fig. 2}
\end{figure}

The \textit{UTA Study Buddy Bot} was built using OpenAI’s API model, deployed via a custom front-end built in Streamlit \citep{baviskar_news_2025, deole_pdf-qa-ai_2024, pokhrel_building_2024}, and integrated in Canvas \citep{sulun_evolution_2018}, allowing students to interact with it through a simple web-based interface. The interface invited students to type their quiz or homework questions and receive guided help, with emphasis on reasoning and scaffolding rather than final answers. An example of the chatbot interface is shown in Fig. \ref{Fig. 2}. The chatbot operated using lightweight retrieval-augmented generation (RAG) techniques, wherein student inputs were matched to a local bank of curated quiz and homework questions \citep{gao_retrieval-augmented_2023, lewis_retrieval-augmented_2020}, and the chatbot then generated step-by-step, Socratic-style explanations. The technical design and implementation of the chatbot is not the focus of this research paper; a companion paper and study will explore the efficacy, design challenges, and broader implications of deploying such AI-based tutors in physics and other university-level courses, along with the student perception(s) of using AI in their classrooms. 

The study addresses the following research questions: (1) What recurring misconceptions and reasoning patterns emerge in student-chatbot conversations in a modern physics course? (2) How can CGT facilitate the identification and interpretation of these patterns across a large textual corpus? (3) What are the broader implications of this approach for AI-driven instruction and PER methodologies?

\subsection{Course Context and Instructional Design}\label{subsec1.2}

The data analyzed in this study were collected from the Fall 2024 semester of PHYS 3313: Modern Physics at the University of Texas at Arlington, taught by Dr. Ramon Lopez. This upper-level undergraduate course was delivered virtual asynchronously, using a flipped-classroom model \citep{bishop_flipped_2013, ozdamli_flipped_2016} adapted for online instruction. Students engaged with course content through a combination of pre-recorded video lectures, assigned textbook readings, and homework problems. The synchronous element traditionally present in flipped classrooms was replaced with collaborative homework discussion opportunities, forums, and AI-powered tools such as the \textit{UTA Study Buddy Bot}. 

The course structure followed 13 instructional modules, with nine content-based chapters covering topics such as Special Relativity, Early Quantum Theory, Quantum Mechanics, Atomic Structure, Molecular and Solid-State Physics, Nuclear Physics, Nuclear Energy, Particle Physics, and Cosmology. The remaining four modules corresponded to three midterm exams and a final comprehensive exam. Each module was aligned with clearly defined Student Learning Objectives (SLOs) \citep{lachlan_student_2012}, which informed both instructional design and assessment. 

One significant pedagogical innovation in this course was the use of AI to generate homework assignments. Rather than relying on conventional problem sets from published textbooks or online databases, the instructor leveraged ChatGPT to generate original, SLO-aligned physics problems tailored to the specific needs and pacing of the class. This approach served multiple functions: it reduced the likelihood of academic dishonesty through online solution-sharing, encouraged deeper engagement with newly presented material, and allowed for rapid iteration and refinement of homework questions in response to emerging student difficulties. A detailed discussion of this generative assessment framework is presented in a separate companion paper \citep{dange_using_2025}. 

This instructional model reflects a broader shift in physics education toward scalable, tech-enhanced learning environments that prioritize active reasoning and student agency \citep{brame_active_2016, prince_does_2004}. The course also built on prior work exploring hybrid and partially flipped formats \citep{yarbrough_student_2024}, extending those insights to a fully asynchronous model. The integration of AI into both content delivery and student support created a data-rich setting for investigating how students interact with modern physics content in real time, particularly when supported by conversational agents designed to mirror tutor-like assistance.

\section{Methodology}\label{sec2}

Using a CGT framework, we analyze a large dataset of student–chatbot conversations collected during a semester-long modern physics course. Our methodological pipeline follows three core phases as discussed earlier in section \ref{subsec1.1}: pattern detection, pattern refinement, and pattern confirmation, each adapted to suit the pedagogical and computational constraints of the dataset. This section outlines the data preprocessing steps and computational tools used to uncover and interpret recurring patterns of student reasoning and misconception. 

The chatbot dataset was collected from conversations between students and the \textit{UTA Study Buddy Bot}, an AI-powered assistant deployed in the Fall 2024 asynchronous modern physics course. Conversations were exported and cleaned to retain only student input. Responses from the bot, timestamps, and system metadata were excluded from analysis. Filtering steps included removing affirmations, short responses under 20 characters, and duplicates to isolate content-rich queries. The result was a corpus of approximately 1500 unique student messages across 13 modules, segmented by instructional chapters and exams. 

Following preprocessing, we adopted a neural topic modeling approach that integrates modern transformer-based embeddings and clustering. While classic topic models like Latent Dirichlet Allocation (LDA) \citep{jelodar_latent_2019} have been widely used, they rely on bag-of-words assumptions and ignore word context, making them less suitable for short, informal student queries \citep{das_gaussian_2015}. In contrast, recent advances in neural topic modeling, particularly the BERTopic framework \citep{grootendorst_bertopic_2022}, offer a more context-aware, modular, and flexible architecture for extracting meaningful themes from language data \citep{egger_topic_2022}. The next section provides an overview of the BERTopic model and its adaptation to our educational dataset.

\subsection{BERTopic}\label{subsec2.1}

BERTopic \citep{grootendorst_bertopic_2022} is a state-of-the-art topic modeling technique that combines transformer-based language models, dimensionality reduction, and density-based clustering to uncover coherent and interpretable topics in text data. It operates in three stages: document embedding, clustering, and topic representation. First, BERTopic generates semantic document embeddings using pre-trained models such as Sentence-BERT (SBERT) \citep{reimers_sentence-bert_2019}, RoBERTa \citep{liu_roberta_2019}, or MiniLM \citep{wang_minilm_2020}. These embeddings encode contextual relationships between words and sentences, allowing similar messages to appear closer in vector space. 

\begin{figure}[h]
\centering
\includegraphics[width=0.8\textwidth]{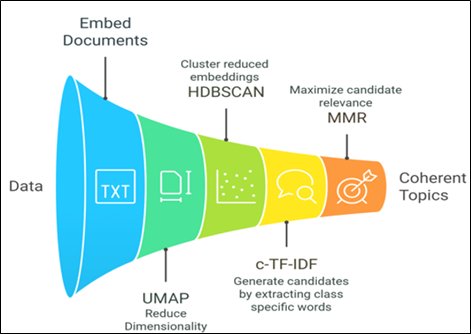}
%\vspace{-1em} % <-- optional if caption is too far
\caption{BERTopic Flowchart for Coherent Topic Clustering}
\label{Fig. 3}
\end{figure}

Once embedded, the high-dimensional vectors are reduced using Uniform Manifold Approximation and Projection (UMAP) \citep{allaoui_considerably_2020, mcinnes_umap_2018}, which preserves both global and local structure. The reduced vectors are then clustered using HDBSCAN \citep{calio_advancing_2022, ibraimoh_comparison_2024}, a soft clustering algorithm that identifies dense regions in the embedding space and automatically classifies noise or outliers. This is particularly useful in educational datasets, where questions can vary widely in specificity and phrasing. Fig. \ref{Fig. 3} describes the entire BERTopic flowchart. 

What sets BERTopic apart is its approach to topic representation. Rather than assuming that topics are defined by words nearest to a cluster centroid, BERTopic uses a class-based TF-IDF (c-TF-IDF) procedure. This method treats each cluster as a single composite document and calculates term frequency–inverse document frequency scores across clusters rather than individual documents. The result is a ranked list of representative words that define each topic in terms of their distinctiveness relative to other topics. This approach has been shown to improve interpretability and coherence in short-text environments \citep{grootendorst_bertopic_2022}. In our study, BERTopic enabled scalable, interpretable topic modeling of student queries across the full semester. Its modularity allowed us to iteratively refine our preprocessing, embedding choice, and clustering parameters while retaining transparency and reproducibility.

\subsection{Case Example: Month of September 2024}\label{subsec2.2}

The first phase of the course, spanning from mid-August through the end of September, encompassed Modules 1 through 4 of the modern physics curricula. These modules introduced students to the Special Theory of Relativity, Early Quantum Theory, and Quantum Mechanics — three foundational yet conceptually challenging chapters. Module 4, which took place during the final week of September, was designated for Exam~1. The content and assessments during this period set the stage for intensive chatbot interaction, as students grappled with unfamiliar frameworks such as time dilation, photon quantization, and wave-particle duality \citep{giancoli_physics_2008}. 

Variations in chatbot engagement throughout September were assessed by monitoring API request volume and token generation. These usage statistics provide insight into when students turned to the \textit{UTA Study Buddy Bot} for help and how their activity patterns corresponded with instructional pacing, homework deadlines, and exam preparation.

\subsubsection{Usage Statistics}\label{subsubsec2.2.1}

\begin{figure}[h]
\centering
\includegraphics[width=\textwidth]{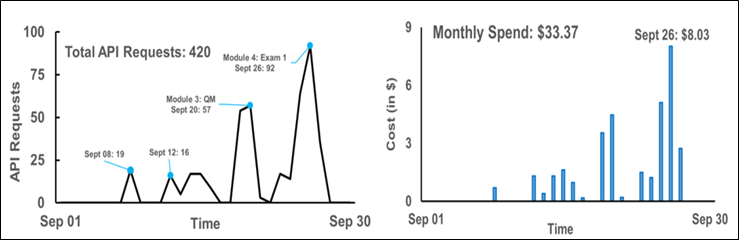}
\vspace{-1em} % <-- optional if caption is too far
\caption{Chatbot Usage and Associated Costs Throughout September}
\label{Fig. 4}
\end{figure}

Fig. \ref{Fig. 4} displays the daily distribution of API requests submitted to the \textit{UTA Study Buddy Bot} during September 2024 - both from a completed conversations and cost perspective. On the left, we can see that usage was initially low at the beginning of the month, with minor peaks on September~8 (19 requests) and September~12 (16 requests), likely tied to early homework assignments. A noticeable rise occurred around September~20, during the Quantum Mechanics module (Module~3), reaching 57 requests. The highest spike occurred on September~26, the day of Exam~1 (Module~4), with 92 recorded interactions, marking a peak in chatbot engagement. 

In the right graph, we show the corresponding monthly expenditure on chatbot API usage. While earlier dates remained under \$3 per day, September~26 saw a sharp surge in cost, reaching \$8.03. This reinforces the observation that student reliance on the bot intensified around assessment periods. In total, chatbot use in September alone resulted in 3.15 million tokens - approximately 108{,}000 words or 245 pages of student-chatbot generated text. 

Together, these metrics reveal how chatbot usage was highly event-driven. Students predominantly sought AI support near deadlines or during moments of conceptual struggle, especially in the days leading up to or coinciding with major exams \citep{marielle_students_2024}. These peaks provide both temporal anchors for interpreting the clusters and thematic context for understanding how misconceptions surface in dialogue.

\subsubsection{Pattern Creation}\label{subsubsec2.2.2}

After processing over 3 million tokens of chatbot interactions, we extracted 344 high-information student messages using a custom Python script tailored for educational dialogue data. This script parsed textual conversation logs, isolated student-authored messages by identifying speaker turns, and applied multiple cleaning steps. These included filtering out short messages (under 20 characters), excluding bot/system responses, and splitting longer inputs into complete, self-contained sentences. It also wrapped Unicode equations and math expressions in standardized LaTeX-style formatting to preserve semantic structure \citep{kopka_guide_2003}. The final dataset retained only semantically rich, natural-language questions that reflected conceptual uncertainty, exam preparation efforts, or problem-solving attempts. These curated inputs were ideal candidates for topic modeling using BERTopic. Examples of retained messages include: 
\begin{itemize} 
\item \textit{``An electron had momentum of 1000 MeV/c, and its rest mass energy is 0.511 MeV (mega-electron volts). What is the kinetic energy of the electron'' 
\item ``How does increasing the width of a potential barrier in the one-dimensional Schrödinger's finite potential well affect the probability of finding the particle on the other side of the barrier?'' 
\item ``Why don't atoms collapse''
\item``If the wave function $\Psi = Ae^{-x^2}$, what is false about it?''}
\end{itemize} 

These actual student questions, drawn directly from the September dataset, demonstrate a mix of precise conceptual queries, incomplete formulations, and domain-specific phrasing. They reflect the language students use when struggling with modern physics concepts, making them highly suitable for topic modeling using BERTopic. The hybrid academic-conversational tone also aligns with the nature of educational dialogue, making them ideal for semantic clustering. The resulting 344-question corpus was embedded, reduced, and clustered using BERTopic as described in earlier sections, forming the foundation for misconception analysis in the following section(s).

\subsubsection{Topic Outputs and Interpretive Insights}\label{subsubsec2.2.3}

BERTopic generated a total of 9 coherent topic clusters from the 344 clean sentences collected during September. Table \ref{tab:bertopic_clusters} shows the topic-wise summary generated by BERTopic from student messages. Each row represents a topic, including the number of messages (``Count''), a compact topic name (``Name''), and the top keywords (``Representation'') that define the semantic core of that cluster. Topic~$-1$ represents the outlier group of messages that did not cluster meaningfully with any dominant theme. This table serves a similar purpose to that presented in another CGT study \citep{tschisgale_integrating_2023}. 

\begin{table*}[htbp]
\caption{Thematic Clusters of Student Reasoning from BERTopic Analysis (September Dataset)}
\label{tab:bertopic_clusters}
\begin{tabularx}{\textwidth}{cl>{\raggedright\arraybackslash}X>{\raggedright\arraybackslash}X>{\raggedright\arraybackslash}X}
\toprule
\textbf{Cluster} & \textbf{Size} & \textbf{Top 5 Words} & \textbf{Sample Sentence} & \textbf{Topic Definition} \\
\midrule
-1 & 33  & N/A & N/A & Outlier/Noise \\
0  & 130 & the, to, and, is, you & What's up fizzbot? & Student Chatbot Socialization \\
1  & 43  & mass, rest, MeV, its, energy & An electron is moving at 0.7c... & Relativistic Energy Confusion \\
2  & 32  & wavelength, electron, with, photon, energy & What is the energy of a photon from n=3 to n=1? & Infinite Square Well Transitions \\
3  & 24  & photon, happens, emitted, 60, energy & A 60 eV photon is emitted when this happens. & Photon Emission Events \\
4  & 24  & harmonic, oscillator, simple, state, drops & A particle in a harmonic oscillator drops from n=4 to n=2. & Harmonic Oscillator State Transitions \\
5  & 18  & state, exited, 10, 1d, in & What is the energy of the first exited state if ground is 10 eV? & Excited States in Oscillators \\
6  & 17  & infinite, square, energy, state, level & What is the fourth excited state if ground is 10 eV? & Square Well Quantization \\
7  & 12  & ground, state, the, what, energy & What is the energy of the ground state? & Ground State Confusion \\
8  & 11  & equation, dimensional, Schrodinger, one, independent & What role does the potential energy term play? & Schrodinger Equation Interpretation \\
\bottomrule
\end{tabularx}
\end{table*}

\underline{Topic 0 (n = 130): Student Chatbot Socialization}\\
Emerged as the largest category, but its content deviated from conventional physics discourse. It included casual, social, and metacognitive language such as \textit{``hmm well I thought the Oppenheimer movie was good''}, \textit{``what’s up fizzbot''}, \textit{``no way i just talked to an AI bot for this long but u cool bro''}, and other personalized musings. These entries revealed students were either giving human traits to the \textit{UTA Study Buddy Bot} or reflecting on big ideas, usually tied to emotional engagement \citep{polyportis_understanding_2025}. The language in this cluster aligned with the chatbot’s peer-like design, suggesting that some students viewed it not just as a tutor, but as a study companion or sounding board during periods of confusion or stress. 

\underline{Topic 1 (n = 43): Relativistic Energy Confusion}\\
This cluster centered on student difficulties with special relativity, especially the distinction between rest mass energy, total energy, and relativistic kinetic energy. Messages commonly included phrases like \textit{``mass MeV rest''} and \textit{``at 0.7c''}, indicating a reliance on formula-based reasoning, but often with conceptual mix-ups. Students frequently misapplied the energy-momentum relation or confused rest energy with kinetic energy, similar to other studies \citep{bryce_momentum_2009, jewett_energy_2008}. 

\underline{Topic 2 (n = 32): Infinite Square Well Transitions}\\
This cluster captured student questions about photon emission during transitions between discrete energy levels in an infinite square well. Representative words like \textit{``wavelength''}, \textit{``photon''}, and \textit{``electron''} appeared frequently, but students often misapplied energy level indices or misunderstood quantum number labeling. 

\underline{Topic 3 (n = 24): Photon Emission Events}\\ Students in this cluster asked short, formulaic questions involving photon emission. Entries often contained numerical constants but little context, suggesting a recall-based approach rather than deep conceptual reasoning. 

\underline{Topic 4 (n = 24): Harmonic Oscillator State Transitions}\\
This topic reflected student confusion with quantum harmonic oscillator systems, especially transitions between different energy states. Phrases such as \textit{``state drops''} and \textit{``simple harmonic oscillator''} were common, but logic regarding energy level spacing or transitions was often unclear.

\underline{Topic 5 (n = 18): Excited States in Oscillators}\\
This cluster centered on early state calculations in harmonic oscillators. Many students used \textit{``exited''} instead of \textit{``excited''}, a likely typo that persisted across the dataset. Despite the spelling, their intent was clear: to calculate energy for low-lying excited states. 

\underline{Topic 6 (n = 17): Square Well Quantization}\\
Students here asked about the progression of energy levels in infinite square wells. Representative words like \textit{``square''}, \textit{``energy''}, and \textit{``state''} suggested mechanical application of quantization formulas rather than a conceptual understanding. 

\underline{Topic 7 (n = 12): Ground State Confusion}\\ This cluster included vague inquiries like \textit{``what is ground state''} without specifying the system. This points to uncertainty about definitions or difficulty connecting terminology to context. 

\underline{Topic 8 (n = 11): Schrödinger Equation Interpretation}\\
Students asked theoretical questions about the one-dimensional Schrödinger equation, including the role of potential energy and the equation’s components. These were often phrased in abstract or conceptual terms.

\begin{figure}[b]
\centering
\includegraphics[width=\textwidth]{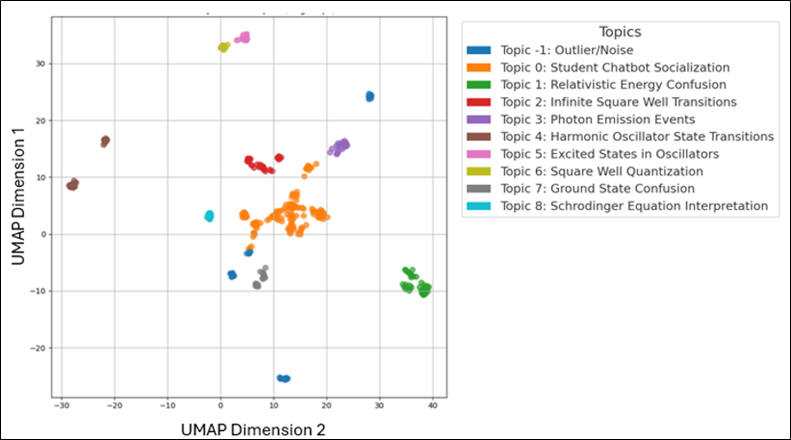}
\vspace{-1em} % <-- optional if caption is too far
\caption{UMAP Projection of Student Interactions (September Dataset)}
\label{Fig. 5}
\end{figure}

\begin{figure}[b]
\centering
\includegraphics[width=\textwidth]{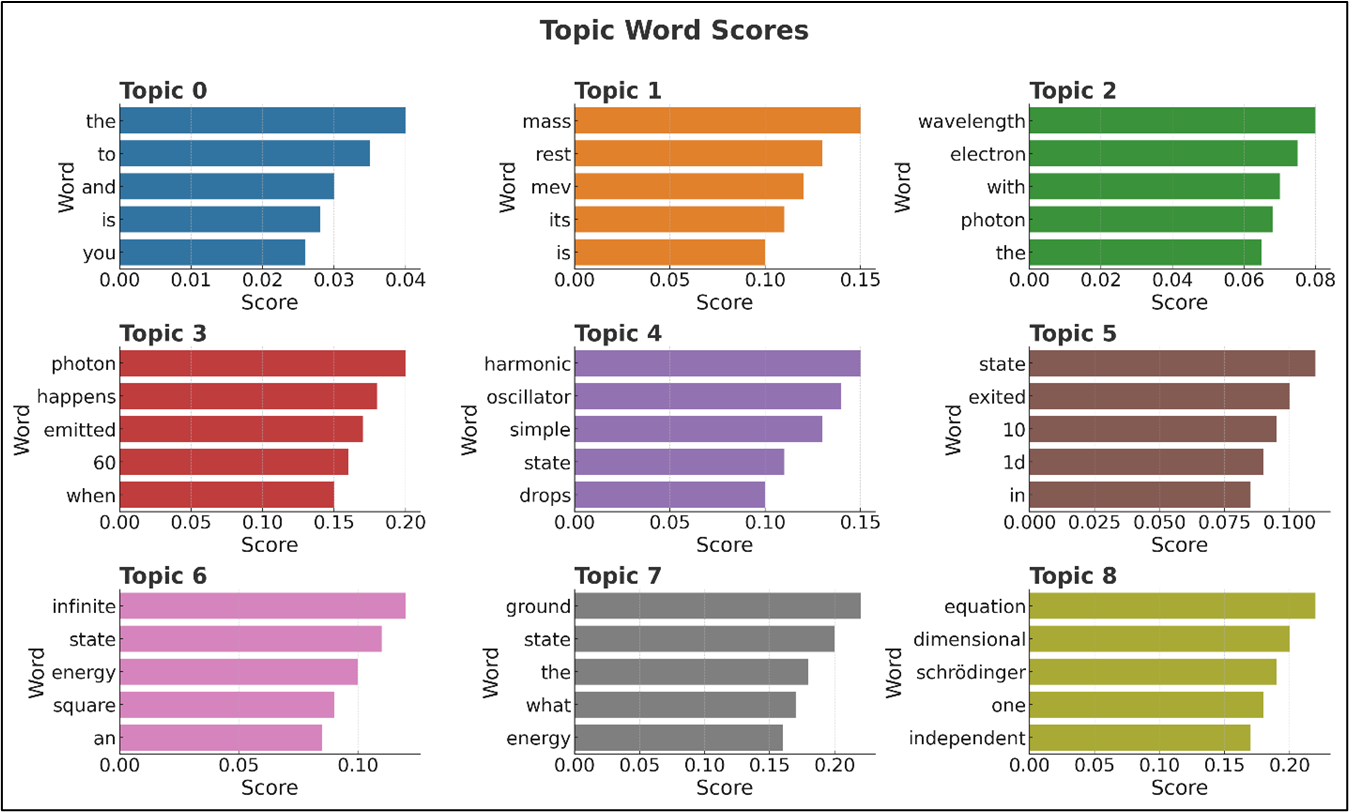}
\vspace{-1em} % <-- optional if caption is too far
\caption{Top Five Distinctive Words per Topic Cluster (September Dataset)}
\label{Fig. 6}
\end{figure}

The following set of two figures (Fig. \ref{Fig. 5} and Fig. \ref{Fig. 6}) provide further insight into the output of BERTopic's modeling of 344 student messages collected during September. These visualizations serve as essential tools for interpreting the types of reasoning, misconceptions, and cognitive frames students brought to their physics inquiries. Fig. \ref{Fig. 5} is a UMAP projection that reduces the high-dimensional sentence embeddings to a two-dimensional space, where each point represents a student-typed utterance and each color denotes a topic assignment. The spatial distribution shows strong internal cohesion and inter-topic separation, validating the semantic distinctiveness of the clusters. The $x$ and $y$-axis labels do not correspond to specific physical variables but rather refer to semantic similarity dimensions: points closer together in this space represent messages that are more similar in meaning. 

For example, Topic 1 (green), associated with relativistic kinetic energy calculations, forms a tight cluster in the lower-right quadrant, while Topic 4 (brown), which deals with transitions in harmonic oscillators, is isolated on the left side. Topic 0 (orange), which contains casual and anthropomorphic interactions with the bot, is centrally located and loosely distributed, reflecting the wide linguistic variance within conversational messages. The presence of Topic $-1$ (dark blue) as an outlier group—spread far from other clusters, confirms BERTopic’s sensitivity to semantically incoherent or fringe messages, such as incomplete thoughts, repeated math symbols, or off-topic entries.

Fig. \ref{Fig. 6} visualizes the five most representative words for each topic cluster, derived using class-based term frequency–inverse document frequency (c-TF-IDF). These terms serve as semantic anchors that help define and label each cluster \citep{alfarizi_emotional_2022, kwon_novel_2024}. The figure shows the normalized c-TF-IDF score, which reflects the relative importance of each word within the topic, versus the actual words. A higher score means the word is more distinctive to that topic. For example:

\begin{itemize} 

\item Topic 1: Words like \textit{mass}, \textit{rest}, \textit{MeV}, and \textit{energy} highlight student focus on relativistic formulas. 

\item Topic 2: Terms such as \textit{wavelength}, \textit{electron}, and \textit{photon} indicate questions involving quantum state transitions in square wells. \item 

Topic 8: Words like \textit{Schrödinger}, \textit{equation}, and \textit{dimensional} confirm that students were engaging with formal theoretical constructs and seeking interpretation rather than calculation. \end{itemize}

Together, these visualizations function as both qualitative and quantitative validation checkpoints. They confirm that the topics discovered by BERTopic reflect pedagogically relevant categories of student thinking, and that the AI model was able to distinguish between computational confusion, theoretical interpretation, and social-emotional dialogue with the chatbot. This multi-modal interpretive lens adds depth to our understanding of how students interact with generative AI tools in physics education contexts.

\section{Semester Wide Analysis}\label{sec3}
This section presents a semester-wide analysis of student–chatbot interactions in the Fall 2024 Modern Physics course. We begin with an exploration of usage statistics to understand when and how frequently students engaged with the \textit{UTA Study Buddy Bot}. We then describe the full-semester implementation of our topic modeling pipeline, including clustering, outlier reassignment, and preparation for macro-level thematic interpretation.

\subsection{Chatbot Usage Statistics Across the Fall 2024 Semester}\label{subsec3.1}

\begin{figure}[h]
\centering
\includegraphics[width=\textwidth]{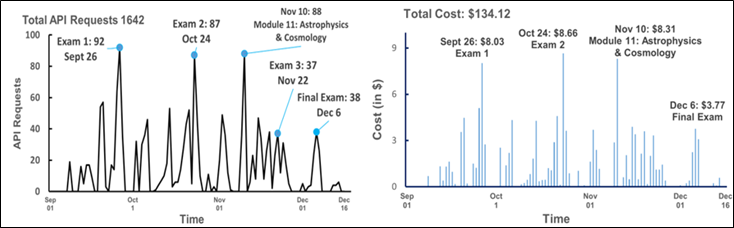}
\vspace{-1em} % <-- optional if caption is too far
\caption{Chatbot Usage and Associated Costs Across the Semester (September – December)}
\label{Fig. 7}
\end{figure}

We begin by examining overall usage trends of the \textit{UTA Study Buddy Bot} throughout the Fall 2024 semester. Fig. \ref{Fig. 7} (left and right panels) visualizes key interaction metrics: the left graph shows the daily number of API requests, while the right graph displays the daily GPT token usage cost in USD. In total, students submitted 1,642 chatbot API requests, generating a semester-long cost of \$134.12. These statistics reflect only the 47 students (out of approximately 62 enrolled) who voluntarily signed the informed consent form and were part of the research study. Students who did not sign the form were still allowed to use an identical chatbot; however, their data was neither logged nor analyzed, in accordance with research ethics protocols. 

Bot engagement here was again, highly event driven. The left graph in Fig. \ref{Fig. 7} reveals distinct usage spikes during Exam~1 on September 26 (92 requests) as noted already from the case study in Section \ref{subsubsec2.2.1}, Exam~2 on October 24 (87 requests), Module~11 on Astrophysics and Cosmology on November 10 (88 requests), Exam~3 on November 22 (37 requests), and the Final Exam on December~6 (38 requests). These spikes strongly align with midterms, the final exam, and a particularly dense module, suggesting that students relied on the chatbot as an on-demand support system during moments of peak academic pressure. 

The right graph in Fig. \ref{Fig. 7} reflects the corresponding GPT token costs. The most expensive days occurred during Exam~2 (\$8.66), Module~11 (\$8.31), and Exam~1 (\$8.03). These high-cost days indicate that students submitted longer or more complex queries that required the model to generate more elaborate responses. Toward the end of the semester, costs slightly decreased despite continued usage, perhaps indicating that students had become more adept at phrasing their questions concisely or knew exactly what type of help to seek. 

The total token count for the semester was 12.654 million input tokens. A token refers to a unit of language processed by GPT to gain semantic meaning: a short word, punctuation mark, or sub-word fragment \citep{grefenstette_what_1994}. While part of this token count came from student messages, a significant portion was generated by the extensive system prompts, instructional scaffolding, and knowledge base context the chatbot accessed each time a student engaged with it. These backend elements were automatically included during every interaction to ensure accurate, helpful responses aligned with the course material. 

Using a conservative estimate, the total input corresponds to roughly 1.5 million words or about 3000 pages of text. Even limiting this estimate to the conversational content between students and the chatbot results in approximately 1,100 pages of raw dialogue. From this large corpus, we extracted 1,504 clean, content-rich student messages for semantic clustering and deeper conceptual analysis. These messages represent focused physics questions, misconceptions, or reasoning attempts and form the empirical basis for the topic modeling process described in the next sub-section. On a per-student basis, this equates to an average of about 35 chatbot requests per student and a GPT usage cost of approximately \$2.85 per student over the entire semester. This relatively low per-user cost highlights the scalability and affordability of AI-based tutoring systems, particularly in large enrollment courses.

\subsection{Topic Modeling across the Fall 2024 Semester}\label{subsec3.2}

After finalizing the preprocessing and refinement protocols on the full dataset of 1,504 student messages, we applied the same BERTopic pipeline from Section \ref{subsec1.1} to generate a complete topic model for the entire semester’s chatbot interaction data. The model initially identified 234 outliers (label $-1$), which was significantly higher than ideal and posed interpretive challenges. Rather than discard these outliers, we performed a second-pass analysis: human review and sub-topic modeling within the outlier pool. This iterative reassignment process successfully reduced the number of outliers to just 18, thereby enhancing the semantic coherence and completeness of the final model. 

The revised model resulted in 47 distinct topic clusters, which were then manually grouped into broader macro level themes described in later sections. Each cluster is labeled with representative keywords and anchored by a sample student message that best captures its theme. While the clustering patterns reflected many of the themes already observed in the case study from September (Section \ref{subsec2.2}), several new topics also emerged, particularly those associated with particle physics, nuclear binding energy, and speculative student inquiry into advanced physics or philosophical topics. 

Although the clustering patterns described above are representative and robust, it is important to acknowledge the stochastic nature of topic modeling algorithms such as BERTopic \citep{egger_topic_2022}. To ensure maximum reproducibility, we deliberately retained all default parameters of BERTopic, including those for HDBSCAN clustering, vectorizer, and topic representation, making our pipeline not only easier to replicate but also aligned with the tool’s intended out-of-the-box usage. We fixed all random seeds (random, numpy, torch, UMAP) at 42 and enforced deterministic behavior, minimizing variability due to hardware or backend execution \citep{ahmed_managing_2022, borcin_optimizing_2024, dutta_seed_2022, schroeder_reliability_2025}.

For the embedding model, we used \texttt{all-MiniLM-L6-v2} \citep{wang_minilm_2020}, a lightweight yet high-performing transformer that balances semantic richness with computational efficiency. This choice ensured consistent and scalable embedding across 1,504 student messages, while still yielding clusters that were interpretable and pedagogically meaningful. Researchers aiming to replicate or extend this study are therefore encouraged to adopt the same embedding model and parameter settings to achieve comparable results. The full codebase, parameter configurations, and topic outputs can be made available by the authors upon request. 

Table \ref{tab:bertopic_semester} below presents a sample of five such clusters, showcasing their distinct top keywords, representative student message, and thematic interpretation. The full list of all 47 clusters and their associated messages is available upon request from the authors. 

While some CGT applications \citep{tschisgale_integrating_2023} ran the full UMAP plus clustering pipeline across a thousand runs to assess the frequency and stability of emergent cluster sizes, our approach took a different but equally grounded path. We prioritized reproducibility and interpretive clarity by retaining all default BERTopic parameters, setting a consistent and constant random seed, and manually refining both initial clusters and outliers. This pipeline produced 47 fine-grained, semantically coherent topics that were stable under 10-fold classification evaluation (see Section \ref{sec4}). Rather than anchoring validity in frequency of occurrence across runs, our model emphasizes thematic coherence, pedagogical relevance, and human-guided reassignment, all of which are core pillars of the pattern refinement stage in CGT. This paper incidentally shows that both methods, frequency-based aggregation and deep interpretive refinement, can provide validity within CGT’s framework and serve different analytical priorities.

\begin{table*}[htbp]
\caption{Semester Wide Thematic Clusters of Student Reasoning from BERTopic Analysis (September - Dec)}
\label{tab:bertopic_semester}
\begin{tabularx}{\textwidth}{ccl>{\raggedright\arraybackslash}X>{\raggedright\arraybackslash}X>{\raggedright\arraybackslash}X}
\toprule
& \textbf{Cluster} & \textbf{Size} & \textbf{Top 5 Words} & \textbf{Sample Sentence} & \textbf{Topic Definition} \\
\midrule

\multirow{14}{*}{\Huge$\left\{\rule{0pt}{8em}\right.$}

\multirow{30}{*}{\rotatebox{90}{\textbf{47 Rows}}}

& -1 & 18  & N/A & N/A & Outlier/Noise \\
& 0  & 87  & energy, bond, binding, potential, ev & Given electron of energy with binding potential… & Potential Energy, Bonding, and Photoelectric Effect \\
& \multicolumn{1}{c}{$\bullet$} & & & & \\
& \multicolumn{1}{c}{$\bullet$} & & & & \\
& \multicolumn{1}{c}{$\bullet$} & & & & \\
& 13 & 41  & Schrodinger, equation, dimension, barrier, potential & How does increasing the width... & Discrete energy levels... \\
& \multicolumn{1}{c}{$\bullet$} & & & & \\
& \multicolumn{1}{c}{$\bullet$} & & & & \\
& \multicolumn{1}{c}{$\bullet$} & & & & \\
& 27 & 24  & reaction, exothermic, mev, endothermic, energy & Which of the following... & Binding Energy and Nuclear Reactions \\
& \multicolumn{1}{c}{$\bullet$} & & & & \\
& \multicolumn{1}{c}{$\bullet$} & & & & \\
& \multicolumn{1}{c}{$\bullet$} & & & & \\
& 46 & 11  & bosons, fermions, behavior, description, examples & I had a question on difference... & Elementary Particles \\
\bottomrule
\end{tabularx}
\end{table*}

\subsection{Aggregating Fine-Grained Topics into macro-themes}\label{subsec3.3}

While the 47 fine-grained clusters identified by BERTopic offered high-resolution insights into student reasoning, this level of detail proves less effective for revealing broader conceptual patterns in student misconceptions. A two-dimensional UMAP projection of all 47 clusters, each with its own color and label, will be visually cluttered and difficult to interpret. In fact, the fine granularity of this view runs counter to the goal of identifying generalizable themes that could inform pedagogical intervention. It then becomes necessary to reduce the dimensionality of interpretation itself by grouping related clusters into higher-order macro-themes \citep{bednarek_balancing_2024}. 

\begin{figure}[h]
\centering
\includegraphics[width=0.9\textwidth]{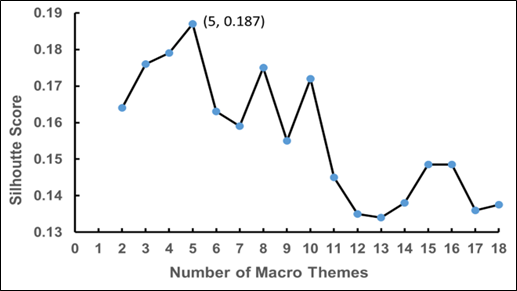}
%\vspace{-1em} % <-- optional if caption is too far
\caption{Silhouette Analysis for Determining Optimal Macro-Theme Count}
\label{Fig. 8}
\end{figure}

To determine how many macro-themes would provide the most meaningful and coherent grouping, we conducted a silhouette analysis using the c-TF-IDF-based vector embeddings of each of the 47 topic centroids. The silhouette score evaluates the internal cohesion of each proposed cluster (how tightly grouped the topics are) relative to the separation from other clusters \citep{pavlopoulos_revisiting_2024, rousseeuw_silhouettes_1987}. This score was calculated for a range of macro-cluster counts ($k = 2$ to $18$), and the result is visualized in Fig. \ref{Fig. 8}. 

The silhouette scores were computed using agglomerative hierarchical clustering with cosine distance as the similarity metric \citep{mullner_modern_2011}. Specifically, we first extracted the centroid-level c-TF-IDF embeddings for all 47 fine-grained topics. These centroid vectors capture the most representative semantic content of each topic. We then applied Agglomerative Clustering across a range of cluster counts ($k = 2$ to $18$), repeatedly grouping the topic centroids into $k$ macro-clusters at each step. For every $k$, we calculated the average silhouette score, which quantifies how similar each topic is to its own cluster compared to others \citep{ogbuabor_clustering_2018}. Higher scores indicate better internal consistency and greater between-cluster separability. This approach, often referred to as silhouette-based cluster validation, enabled us to determine that $k = 5$ provided the most interpretable and coherent macro-theme structure for our dataset. 

The $x$-axis was intentionally capped at 18. This upper bound was not arbitrary, but rather pedagogically grounded. The course content was organized into 9 instructional modules, excluding exams. If we conservatively estimate that each module contains at most two distinct conceptual domains where student misconceptions tend to cluster, then 18 becomes a natural ceiling for the number of macro-themes that can be meaningfully supported by the curriculum. Moreover, allowing more than 18 themes would risk reintroducing the very granularity we aimed to reduce, blurring the line between macro-level interpretation and fine-grained clustering. Thus, limiting the analysis to 18 candidate themes strikes a balance between theoretical and curricular realism.

\subsection{Visualizing and Interpreting macro-themes}\label{subsec3.4}

\begin{figure}[h]
\centering
\includegraphics[width=\textwidth]{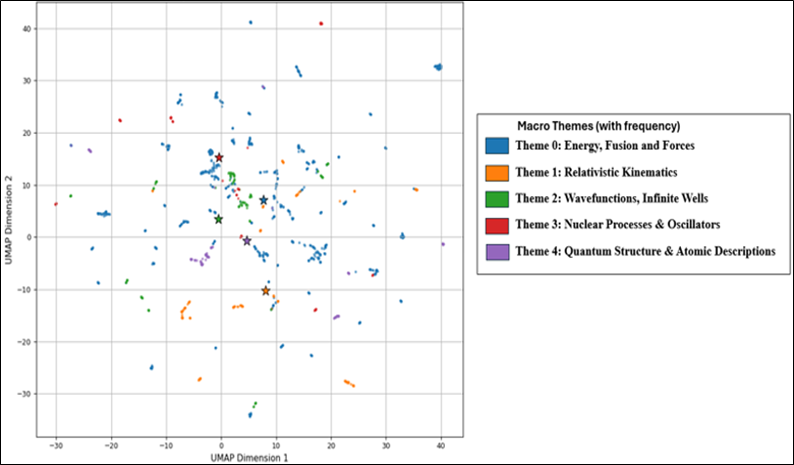}
\vspace{-1em} % <-- optional if caption is too far
\caption{UMAP Projection of 1486 Student Messages Colored by Macro Theme}
\label{Fig. 9}
\end{figure}

To qualitatively validate the results of our macro-theme clustering, we projected all student messages into a two-dimensional UMAP embedding space and colored them according to their assigned macro-themes. Fig. \ref{Fig. 9} shows this visualization, where each dot represents an individual student message and each large star marks the centroid of a macro-theme. The five macro-themes are distinguished by color and labeled in the legend according to their frequency in the dataset. This layout offers a visual overview of where different forms of student reasoning are situated within the latent semantic space.

These macro-themes reflect the most common domains in which student reasoning clustered as follows:

\begin{itemize} 

\item \textbf{Macro-Theme 0: Energy, Fusion, and Forces in Modern Physics}\\ 
This dominant theme includes student messages about a wide range of concepts spanning nuclear fusion, fundamental forces, binding energy, and astrophysical processes.

\item \textbf{Macro-Theme 1: Relativistic Kinematics}\\
Students in this cluster struggled with relativistic motion, rest mass energy, and kinetic energy concepts, often asking questions involving MeV, particle behavior, and the differences between classical and relativistic formulations. 

\item \textbf{Macro-Theme 2: Wavefunctions \& Infinite Wells}\\ This theme combines technical quantum mechanics content such as square wells and state transitions with early quantum theory. 

\item \textbf{Macro-Theme 3: Nuclear Processes \& Oscillators}\\
This theme contains student thinking around beta decay, half-life calculations, and harmonic oscillators. Students often expressed confusion in applying exponential decay equations or reasoning about quantized transitions in oscillator systems. 

\item \textbf{Macro-Theme 4: Quantum Structure and Atomic Descriptions}\\ 
This theme focuses on how students describe and conceptualize atomic structure, including orbitals, quantum numbers, and measurement ideas. It reflects misunderstanding of how quantization and spatial properties relate in atomic and subatomic systems. 

\end{itemize} 

Macro-theme~0 (the first theme) represents the single largest category in the dataset, capturing 974 of the 1,486 student messages—approximately 65\% of the total. It encompasses a wide array of topics: energy, fusion and fission, fundamental forces, solar processes, and energy transfer across systems. The dominance of Theme~0 likely stems from topic imbalance at the fine-grained level, where multiple large topics were already centered on energy-related ideas. During macro clustering, these topics, rich in general-purpose physics terms like \textit{``energy''}, \textit{``force''}, and \textit{``mass''} pulled semantically similar clusters into a single, expansive category. This effect was reinforced by cosine-based agglomerative clustering, which naturally merges overlapping content, and possibly by choosing a relatively low number of macro-themes ($k = 5$). While this outcome reflects the central role of \textit{``energy''} in physics, it introduces an interpretive imbalance that future work could address by further sub-clustering Theme~0 or experimenting with higher values of $k$ to improve thematic granularity. To complement this visualization, Table \ref{tab:macro_theme_summary} presents a summary of the five macro-themes, including their cluster ID, size, top five TF-IDF words, and descriptive titles based on an interpretive synthesis of their contents. The analysis in this section is based on the final set of 1,486 valid student messages, after the removal of 18 un-reassignable outliers.

\begin{table*}[htbp]
\caption{Macro Theme Summary}
\label{tab:macro_theme_summary}
\begin{tabularx}{\textwidth}{cl>{\raggedright\arraybackslash}X>{\raggedright\arraybackslash}X}
\toprule
\textbf{Macro Cluster} & \textbf{Size} & \textbf{Top 5 Words} & \textbf{Macro Theme} \\
\midrule
0 & 974  & energy, force, fusion, mass, answer & Energy, Fusion \& Forces \\
1 & 170  & rest, kinetic, mass, photon, wavelength & Relativistic Kinematics \\
2 & 144  & Square, state, infinite, electron, ground & Wavefunctions \& Infinite Wells \\
3 & 85   & decay, harmonic, beta, simple, continuously & Nuclear Processes \& Oscillators \\
4 & 113  & orbital, number, quantum, parallax, arcseconds & Quantum Structure and Atomic Descriptions \\
\bottomrule
\end{tabularx}
\end{table*}

\section{Pattern Confirmation and Accuracy Analysis}\label{sec4}

To validate the conceptual and statistical reliability of the five macro-themes identified through unsupervised clustering, we implemented a supervised classification analysis. Our goal was to ensure that these high-level groupings were not only conceptually sound but also structurally consistent enough to be predicted from sentence-level embeddings. To do this, we trained a classifier to assign each sentence to one of five macro-themes and evaluated its performance using 10-fold cross-validation, a widely used method for estimating out-of-sample accuracy \citep{Sreedharan_leave-one-out_2023}. In 10-fold cross-validation, the dataset is randomly partitioned into ten equally sized chunks or ``folds.'' For each of the ten iterations, nine folds are used for training and the remaining one is used for testing. This process is repeated so that each sentence serves as a test case exactly once. The accuracy scores from all ten iterations are then averaged to obtain a more reliable performance estimate. This method is especially useful for medium-sized datasets like ours ($\sim$1500 sentences), as it maximizes the use of all data while mitigating risks of overfitting or performance instability due to an unlucky train-test split \citep{kohavi_study_1995}. 

Our classifier achieved an average accuracy of $0.90 \pm 0.02$, meaning that it correctly predicted the macro-theme for 90\% of the sentences, with a modest deviation of 2\%. This level of performance suggests that the macro-themes are not arbitrary: they reflect coherent patterns that are statistically learnable from sentence embeddings. The low variation across folds underscores the stability and robustness of the classification pipeline. While some misclassifications did occur, they tended to fall near conceptual boundaries, such as between ``Energy, Fusion \& Forces (Theme 0)'' and ``Wavefunctions \& Infinite Wells (Theme 2),'' where student language naturally overlaps. This is an expected and even useful source of noise, indicating areas of instructional ambiguity or conceptual blending. 

While studies like \citet{tschisgale_integrating_2023} used a Relevance Vector Machine (RVM) \citep{tzikas_tutorial_2006} to validate their macro-themes, we employed logistic regression for its simplicity, interpretability, and strong performance on high-dimensional sentence embeddings. Logistic regression (LR) offers probabilistic outputs, integrates well with standard machine learning libraries, and delivers stable results with minimal computational overhead. In our case, it served the same purpose as RVM, testing whether macro-themes are statistically learnable, while offering greater ease of implementation and reproducibility \citep{krishnapuram_applying_2002}. In many medical applications, LR’s superiority is attributed to ``simplicity and interpretability'' \citep{arshad_power_2023}, and studies show that LR can achieve ``comparable performance to RVM'' \citep{zanon_sparse_2020}. Future work can extend this analysis by comparing logistic regression with alternative classifiers to identify the most effective model for accuracy, robustness, and instructional insight. 

\begin{figure}[h]
\centering
\includegraphics[width=0.7\textwidth]{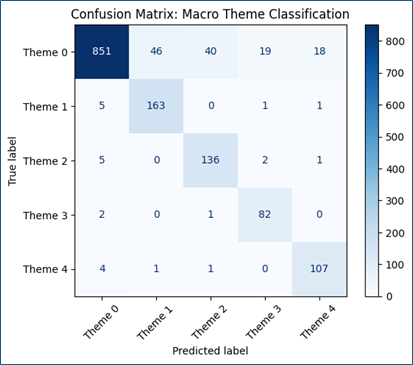}
%\vspace{-1em} % <-- optional if caption is too far
\caption{Confusion Matrix for Macro Theme Classification Using 10-Fold Cross-Validation}
\label{Fig. 10}
\end{figure}

To further investigate classifier performance, we visualized the results using a confusion matrix (see Fig. \ref{Fig. 10}). The confusion matrix compares true macro-theme labels (y-axis) with the predicted labels (x-axis), where diagonal values represent correct classifications. Higher values along the diagonal indicate stronger model accuracy, while off-diagonal values show where the model confused similar themes. A strong confusion matrix exhibits a dominant diagonal, as seen in our case, with values such as 851 (out of 974) correctly classified in ``Energy, Fusion \& Forces'' and 163 (out of 170) in ``Relativistic Kinematics,'' indicating the model’s ability to reliably distinguish between major conceptual themes. This visualization helps readers assess the classifier’s reliability and understand where student thinking overlaps between related concepts. 

Finally, in addition to these quantitative checks, we conducted a round of human inspection to assess thematic coherence. A sample of sentences from each macro-theme was reviewed by the authors familiar with both physics content and student reasoning. This inspection confirmed that the macro-themes were conceptually meaningful and pedagogically distinct. Importantly, we found no clear-cut errors that warranted human correction; the only instances of disagreement arose in genuinely ambiguous cases, such as sentences that could reasonably belong to multiple themes (e.g., both ``Quantum Structure'' and ``Wavefunctions''). In these legitimate grey areas, we opted to retain the model’s original assignments, recognizing that human judgment is inherently subjective while the model applies its criteria consistently. Overall, the model performed robustly, and our review affirmed its reliability without much need for manual intervention.

\section{Discussion and Significance}\label{sec5}

The use of a programmable, deployable AI-powered chatbot as a research probe in physics education is a novel methodological contribution of this study. Unlike traditional tutoring systems or data collection tools \citep{vanlehn_relative_2011}, the chatbot in this study (custom-built and embedded into the instructional infrastructure of a real modern physics course) served both as a support system and a data source. Its conversational design enabled students to articulate their reasoning, confusions, and problem-solving strategies in an informal, peer-like setting, outside of instructor prompts or test conditions. This configuration created a continuous, low-friction interface for capturing authentic student thought processes at scale. 

The novelty lies not just in deploying a chatbot, but in how it was systematically analyzed using the principles of CGT. This study demonstrates that once student-chatbot dialogue is collected, it can be mined using scalable, reproducible computational methods, without sacrificing the interpretive depth that qualitative PER demands. The CGT framework allowed us to identify recurring patterns of conceptual difficulty, reorganize them through human-guided refinement, and validate them through semantic clustering and classification. This is a fundamentally new workflow for physics education research: a scalable, low-cost, and ethically sound way to access thousands of moments of conceptual struggle, framed in the students’ own language. 

What makes this pipeline powerful is its modularity and ease of implementation. The chatbot itself is inexpensive to run (less than \$10 per student over a full semester), and its data can be processed with open-source NLPs, ML and clustering tools. Once designed, this system can be replicated across classrooms, institutions, and even disciplines. This presents a scalable blueprint for integrating AI-based educational tools not just as instructional supports, but as research instruments capable of capturing high-volume qualitative data in a structured, analyzable format. 

\begin{figure}[h]
\centering
\includegraphics[width=\textwidth]{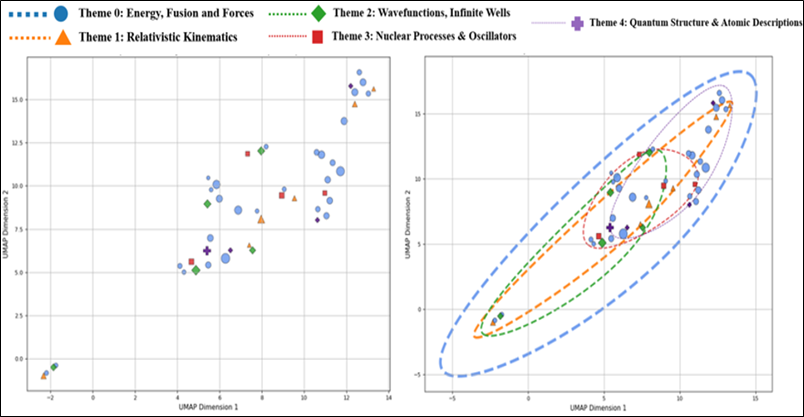}
\vspace{-1em} % <-- optional if caption is too far
\caption{UMAP Projection of 47 Topic Centroids Without and With Thematic Enclosure Lines}
\label{Fig. 11}
\end{figure}

The value of this approach is especially evident in the macro-level clustering and visualization of student reasoning. Fig. \ref{Fig. 11} shows how 1,500+ student messages were semantically organized into five emergent conceptual zones. These visualizations reveal the shape and boundaries of student understanding, with centroid layouts and smoothed enclosures marking coherent zones of reasoning. For instance, the semantic overlap between clusters in ``Wavefunctions \& Infinite Wells'' and ``Quantum Structure'' signals a persistent area of blending or confusion---a place where instruction might benefit from clearer conceptual demarcation or reinforced scaffolding. These representations transform abstract NLP outputs into pedagogically actionable insights. 

Each point in the plot represents the centroid of a single fine-grained topic, essentially the average position in UMAP space of all student messages within that topic. The size of each centroid marker (and the corresponding thickness of the thematic lines) corresponds to the number of student messages (i.e., sentence count) contained within that topic, offering a visual proxy for how common or prevalent that theme was in the semester’s conversations. The right-hand side plot builds on this insight with concentric hull enclosures that trace the shape and semantic spread of each macro theme. These dashed contours are ordered from the innermost and smallest to the outermost and broadest, visually encoding how themes grow in conceptual breadth and student volume. 

The thinnest enclosures (e.g., those surrounding Quantum Structure or Nuclear Processes) represent tightly defined, specialized zones of inquiry---often linked to specific course chapters or well-scaffolded learning objectives. In contrast, the largest and outermost contour, corresponding to Theme 0: Energy, Fusion \& Forces, encloses the greatest number of topic centroids and covers a wide semantic area. The concentric hulls show how student misconceptions evolve from focused topics to broader themes, with the final outer hull enclosing Energy, Fusion \& Forces representing the thematic convergence point for many threads of modern physics instruction, and simultaneously, the area where instructional attention must be most carefully focused. 

From a PER perspective, this work reframes what counts as ``qualitative data.'' Instead of being limited to a dozen interviews or hand-coded short answers, we now have access to full-semester conversational logs containing real-time traces of learning. When analyzed through a CGT lens, this data enables educators to move beyond anecdotal misconceptions and instead chart structured, thematic maps of student understanding. This is especially crucial in domains like quantum mechanics and relativity \citep{leonardi_analysis_2024, styer_common_1996}, where student intuition is often mismatched with formal models and where surface-level correctness may mask deeper conceptual gaps \citep{mcdermott_guest_1993}.

\section{Conclusion}\label{sec6}

In conclusion, this study presents a novel, scalable methodology for capturing and analyzing student reasoning in physics education by embedding an AI-powered chatbot into an asynchronous course and applying Computational Grounded Theory (CGT) to the resulting dialogue data. Across five key sections, we described how student-chatbot conversations were collected, preprocessed, clustered into fine-grained topics, aggregated into macro themes, and validated through both supervised classification and human interpretation. The findings reveal persistent misconceptions in quantum and relativistic physics while also demonstrating how natural language modeling can surface the conceptual contours of student thinking in ways previously unimaginable at this scale. By combining open-source NLP frameworks like BERTopic with pedagogically grounded research questions, we offer a replicable pipeline that transforms messy, real-world student data into structured insights. This work not only exemplifies the future of large-scale qualitative research in education but also opens the door to cross-disciplinary applications in chemistry, biology, or humanities, where AI tools and CGT can together generate actionable understanding. As open-source alternatives to proprietary models like OpenAI continue to evolve \citep{abnar_llama_2024, geminiteamGemini15Unlocking2024}, future explorations may build similarly powerful pipelines that are more transparent, accessible, and adaptable to different domains, paving the way for democratized and data-driven education research across diverse contexts.

The findings of this study address the three guiding research questions posed in Section \ref{subsec1.1}. (1) The application of BERTopic revealed a range of recurring misconceptions and reasoning patterns in student chatbot conversations, such as confusion between relativistic kinetic energy and rest mass energy, misunderstanding of quantum energy level transitions, and vague references to ground states and equations. These misconceptions were not only persistent but also surfaced in the students’ own language, allowing for a more nuanced understanding of how learners articulate uncertainty, apply formulas, or interpret abstract ideas in modern physics. 

(2) CGT facilitated the identification and interpretation of these patterns through a scalable, replicable pipeline that combined semantic embeddings, unsupervised clustering, and human-in-the-loop interpretation. This allowed us to move from over 1,500 fragmented queries to five coherent macro themes of reasoning. In doing so, CGT not only made large-scale student discourse analyzable but also demonstrated how AI and qualitative methods can converge to reveal the conceptual architecture of student thinking in physics.

Lastly, (3) the broader implication of this approach is its potential to transform AI-powered tools into research instruments, ones that can uncover meaningful learning patterns without relying on pre-structured diagnostics. By capturing learning into pedagogically relevant categories, CGT advances the methodological toolkit of physics education research and provides a roadmap for future studies at the intersection of AI and instruction. Used in combination with traditional qualitative methods like interviews and surveys, this approach can significantly reduce the time, labor, and cost required to extract meaning from large-scale student discourse, making it a practical and scalable tool for education research.

\backmatter

\bmhead{Acknowledgements}
We would like to thank Aniket Didolkar for his introduction to BERTopic and for his assistance in interpreting the clustering data.

\section*{Declarations}

\bmhead{Conflict of Interest}
All authors certify that they have no affiliations with or involvement in any organization or entity with any financial interest or non-financial interest in the subject matter or materials discussed in this manuscript.

\bmhead{Funding}
This work was supported in part by a Teaching Innovation Research grant at The Center for Research on Teaching and Learning Excellence (CRTLE) at the University Texas at Arlington.

\bmhead{Data Availability}
Data sets generated during the current study are available from the corresponding author on reasonable request. Institutional Review Board (IRB)  restrictions apply to the availability of these data, and so are not publicly available.

\bmhead{Author Contributions}
All authors contributed to the study’s conception and design. Material preparation, data collection, and analysis were performed by Atharva Dange. The first draft of the manuscript was written by all authors who commented on previous versions. All authors read and approved the final manuscript.

%%%%%%BIBLIOGRAPHY .bbl%%%%%%%%%%%%%%%%%%%%%%
%% BioMed_Central_Bib_Style_v1.01
% BibTex style file: bmc-mathphys.bst (version 2.1), 2014-07-24
\ifx \bisbn   \undefined \def \bisbn  #1{ISBN #1}\fi
\ifx \binits  \undefined \def \binits#1{#1}\fi
\ifx \bauthor  \undefined \def \bauthor#1{#1}\fi
\ifx \batitle  \undefined \def \batitle#1{#1}\fi
\ifx \bjtitle  \undefined \def \bjtitle#1{#1}\fi
\ifx \bvolume  \undefined \def \bvolume#1{\textbf{#1}}\fi
\ifx \byear  \undefined \def \byear#1{#1}\fi
\ifx \bissue  \undefined \def \bissue#1{#1}\fi
\ifx \bfpage  \undefined \def \bfpage#1{#1}\fi
\ifx \blpage  \undefined \def \blpage #1{#1}\fi
\ifx \burl  \undefined \def \burl#1{\textsf{#1}}\fi
\ifx \doiurl  \undefined \def \doiurl#1{\url{https://doi.org/#1}}\fi
\ifx \betal  \undefined \def \betal{\textit{et al.}}\fi
\ifx \binstitute  \undefined \def \binstitute#1{#1}\fi
\ifx \binstitutionaled  \undefined \def \binstitutionaled#1{#1}\fi
\ifx \bctitle  \undefined \def \bctitle#1{#1}\fi
\ifx \beditor  \undefined \def \beditor#1{#1}\fi
\ifx \bpublisher  \undefined \def \bpublisher#1{#1}\fi
\ifx \bbtitle  \undefined \def \bbtitle#1{#1}\fi
\ifx \bedition  \undefined \def \bedition#1{#1}\fi
\ifx \bseriesno  \undefined \def \bseriesno#1{#1}\fi
\ifx \blocation  \undefined \def \blocation#1{#1}\fi
\ifx \bsertitle  \undefined \def \bsertitle#1{#1}\fi
\ifx \bsnm \undefined \def \bsnm#1{#1}\fi
\ifx \bsuffix \undefined \def \bsuffix#1{#1}\fi
\ifx \bparticle \undefined \def \bparticle#1{#1}\fi
\ifx \barticle \undefined \def \barticle#1{#1}\fi
\bibcommenthead
\ifx \bconfdate \undefined \def \bconfdate #1{#1}\fi
\ifx \botherref \undefined \def \botherref #1{#1}\fi

%\ifx \url \undefined \def \url#1{\textsf{#1}}\fi
\providecommand{\url}[1]{\textsf{#1}}

\ifx \bchapter \undefined \def \bchapter#1{#1}\fi
\ifx \bbook \undefined \def \bbook#1{#1}\fi
\ifx \bcomment \undefined \def \bcomment#1{#1}\fi
\ifx \oauthor \undefined \def \oauthor#1{#1}\fi
\ifx \citeauthoryear \undefined \def \citeauthoryear#1{#1}\fi
\ifx \endbibitem  \undefined \def \endbibitem {}\fi
\ifx \bconflocation  \undefined \def \bconflocation#1{#1}\fi
\ifx \arxivurl  \undefined \def \arxivurl#1{\textsf{#1}}\fi
\csname PreBibitemsHook\endcsname

%%%%%%%%%%BIBLIOGRAPHY BBL%%%%%%%%%%%%%%%%

\end{document}